\documentclass{aa}  

\newcommand{\beq}[1]{\begin{equation}\label{#1}}
\newcommand{\eeq}{\end{equation}}
\newcommand{\sub}[1]{_{\rm #1}}

\newcommand{\rhos}{\rho\sub{\star}}
\newcommand{\rhop}{\rho\sub{p}}
\newcommand{\Rs}{R\sub{\star}}
\newcommand{\dR}{d\sub{R}}
\newcommand{\AU}{\mathrm{au}}

\usepackage{graphicx}
\usepackage{txfonts}
\usepackage{natbib}
\usepackage{orcidlink}
\usepackage{hyperref}
\usepackage{amsmath}
\usepackage{enumitem}
\setlist[itemize]{label=--}

\hypersetup{
  colorlinks,
  citecolor=cyan,
  linkcolor=magenta,
  urlcolor=teal,
}
\bibpunct{(}{)}{;}{a}{}{,} 

\begin{document} 
 
   \title{Observational Signatures of Planetary Tidal Disruption Events Around Solar-Mass Stars}
\author{
    Matías Montesinos\inst{1}\thanks{E-mail: \href{mailto:matias.montesinosa@usm.cl}{matias.montesinosa@usm.cl}}\orcidlink{0000-0001-9789-5098}
    \and
    Sergei Nayakshin\inst{2}\orcidlink{0000-0002-6166-2206}
    \and
    Vardan Elbakyan\inst{3,4}\orcidlink{0000-0002-9433-900X}
    \and
    Zhen Guo\inst{5,6}\orcidlink{0000-0003-0292-4832}
     \and
    Mario Sucerquia\inst{7}\orcidlink{0000-0002-8065-4199}
    \and
    Amelia Bayo\inst{8}\orcidlink{0000-0001-7868-7031}
    \and
    Zhaohuan Zhu\inst{9,10}\orcidlink{0000-0003-3616-6822}
}

\institute{
    Departamento de Física, Universidad Técnica Federico Santa María, Avenida España 1680, Valparaíso, Chile
    \and
    School of Physics and Astronomy, University of Leicester, Leicester LE1 7RH, UK
    \and
    Fakult\"at f\"ur Physik, Universit\"at Duisburg-Essen, Lotharstra\ss e 1, D-47057 Duisburg, Germany
    \and
    Research Institute of Physics, Southern Federal University, Rostov-on-Don 344090, Russia
    \and
    Instituto de Física y Astronomía, Universidad de Valparaíso, ave. Gran Bretaña, 1111, Casilla 5030, Valparaíso, Chile
    \and
    Millennium Institute of Astrophysics, Nuncio Monseñor Sotero
    Sanz 100, Of. 104, Providencia, Santiago, Chile
    \and
    Univ. Grenoble Alpes, CNRS, IPAG, 38000 Grenoble, France
    \and
    European Southern Observatory, Karl-Schwarzschild-Strasse 2, 85748 Garching bei M\"unchen, Germany
    \and 
    Department of Physics and Astronomy, University of Nevada, Las Vegas, 4505 S. Maryland Pkwy, Las Vegas, NV 89154, USA
    \and
    Nevada Center for Astrophysics, University of Nevada, Las Vegas, 4505 S. Maryland Pkwy., Las Vegas, NV 89154-4002, USA
}

   \date{Received August 8, 2025; accepted TBD}

\abstract
{The tidal disruption of planets by their host stars represents a growing area of interest in transient astronomy, offering insights into the final stages of planetary system evolution and the scattering of planets in gas-poor environments.}
{We aim to model the hydrodynamic evolution and predict the multi-wavelength observational signatures of planetary TDEs around a solar-mass host, focusing on Jupiter-like and Neptune-like progenitors and examining how different eccentricities of the planet’s pre-disruption orbit shape the morphology and emission of the tidal debris.}
{We perform 2D hydrodynamic simulations using the \texttt{FARGO3D} code to model the formation and viscous evolution of the resulting debris disk. We employ a viscous $\alpha$-disk prescription and include a time-dependent energy equation to compute the disk's effective temperature and subsequently derive the bolometric and multi-band photometric light curves.}
{Our simulations show that planetary TDEs produce a diverse range of luminous transients. A Jupiter-like planet disrupted from a circular orbit at the Roche limit generates a transient peaking at $L_{\rm bol} \sim 10^{38}$~erg~s$^{-1}$ after a $\sim$12-day rise. In contrast, the same planet on an eccentric orbit ($e=0.5$) produces a transient of comparable peak luminosity but on a much shorter timescale, peaking in only $\sim$1~day and followed by a highly volatile light curve. We find that the effect of eccentricity is not universal, as it accelerates the event for Jupiter but delays it for Neptune. A robust "bluer-when-brighter" colour evolution is a common feature as the disk cools over its multi-year lifetime.}
{The strong dependence of light curve morphology on the initial orbit and progenitor mass makes these events powerful diagnostics. The dichotomous effect of eccentricity indicates light curves probe both orbital parameters and the planet's internal structure. This framework is crucial for identifying planetary TDEs in time-domain surveys.}


   \maketitle

\section{Introduction}

Tidal Disruption Events (TDEs) are astrophysical phenomena in which an astronomical object is gravitationally disrupted upon closely approaching a more massive object. Historically, TDEs were first described theoretically to explain the disruption of stars by supermassive black holes (SMBHs) at galactic centers. Early analytical models by \citet{Rees1988} laid the foundational framework describing how tidal forces dominate a star's self-gravity, leading to its disruption, the subsequent accretion of debris onto the black hole, and observable transient phenomena.

The observational confirmation of stellar TDEs involving SMBHs has been supported by the discovery of luminous flares across the electromagnetic spectrum \citep[for a review, see][]{Gezari2021}. A well-known example is Swift J1644+57, interpreted as the relativistic disruption of a star by a dormant SMBH \citep{Bloom+2011}, while other events have shown diverse multi-wavelength signatures \citep{Elbakyan+2023}. Complementing these observations, numerical simulations have become crucial for interpreting the data. For instance, \citet{Montesinos+Pacheco2011} presented one of the earliest hydrodynamic simulations of the accretion phase following a stellar TDE, modeling the formation and evolution of the accretion disk. Later, \citet{Montesinos+Pacheco2013} applied these models to interpret the photometric properties of PS1-10jh, a TDE candidate with unusually low disk temperatures, presenting a theoretical model for its light curve. Such work has helped confirm the existence of SMBH-related TDEs, providing crucial tests for theoretical models. Beyond total disruptions, recent discoveries have revealed related phenomena such as Quasi-Periodic Eruptions (QPEs), which may involve repeating partial disruptions of stellar or substellar companions \citep{Hernandez+2025}.

In recent years, the study of TDEs has expanded to include interactions involving stellar-mass black holes, neutron stars, and white dwarfs, with theoretical rates still being actively investigated \citep[e.g.,][]{Stone+2019}. For stellar-mass black holes, models indicate that these lower-mass compact objects can tidally disrupt stars in dense environments like globular clusters, producing transient signals in X-rays and gamma rays \citep[e.g.,][]{Kremer2019b}.

The physics of tidal disruption also extends to planetary-mass objects. In dense stellar environments, dynamical interactions can produce a population of free-floating planets. Simulations suggest that these liberated planets can subsequently experience TDEs by stellar-mass black holes, neutron stars, or white dwarfs at significant rates \citep{Kremer+2019}. Specific scenarios, such as the partial disruption of rocky planets by neutron stars, could produce observable fragments \citep{Kurban+2023}. Furthermore, the TDE of planets or planetesimals by white dwarfs is a leading explanation for the observed metal pollution in their atmospheres \citep{Koester+2014}.

While the disruption of planetary bodies by compact objects has been extensively studied, dynamical instabilities can also drive planets into catastrophic close encounters with their main-sequence host stars \citep{Metzger+2012}. Previous numerical efforts, employing both smoothed particle hydrodynamics (SPH) and grid-based methods, have explored the dynamics of planetary disruptions primarily in parabolic or highly eccentric trajectories around various types of hosts \citep[e.g.,][]{Faber+2005, Guillochon+2011, Liu+2013}. 

These planetary tidal disruption events are predicted to generate a distinct class of optical and X-ray transients powered by the accretion of the planetary material. Regardless of the specific stellar environment, the sudden injection of high-density planetary debris generates a luminous accretion flow that typically overwhelms the pre-existing emission of the host system. Modeling this process is challenging, as the evolution of the resulting debris is governed by complex hydrodynamics, shock heating, and viscous transport, which together dictate the observational signatures \citep{Zubovas+Nayakshin2012, Nayakshin+2015}.

Complementary to these multi-dimensional efforts, 1D hydrodynamical studies have been crucial in determining the stability of the disruption process. \citet{Nayakshin+Lodato2012} showed that Roche Lobe OverFlow (RLOF) can lead to either stable mass transfer or prompt disruption. More recently, \citet{Nayakshin+Elbakyan2024} incorporated detailed planetary structure models to demonstrate that the outcome is governed by the planet's radial response to mass loss: if the planet contracts, the system settles into a long-term steady state, whereas if it expands, the mass loss accelerates, culminating in a prompt catastrophic disruption.

A natural pathway towards these planetary TDEs is the inward migration and orbital decay of ultra-short-period planets (USP): as shown by studies of late-stage inspiral (e.g., \citealt{Jackson2016}; \citealt{Spina2024}; \citealt{Alvarado-Montes2025}), close-in giants can gradually lose angular momentum until they cross their Roche limit, making tidal disruption by the host star an unavoidable end state for part of the USP population.

Recent studies have also explored the secondary consequences of planetary disruption. In particular, \citet{Sucerquia+Montesinos2025} investigated the fate of volatile-rich gas released during the tidal disruption of giant planets. Their work showed that this material could be captured by outer planets in the same system, forming transient secondary atmospheres and potentially explaining the volatile content of some exoplanets.

This theoretical framework has become increasingly relevant as observational evidence for stars consuming their planets is now emerging. The optical transient ZTF~SLRN-2020 was interpreted as the first real-time observation of a star engulfing a planet in a direct merger \citep{De+2023}, an event resembling a scaled-down luminous red nova. Other work has found evidence of past consumption events from chemical signatures in stellar atmospheres \citep{Lau+2025}. More recently, significant variability observed in the young star VVV-WIT-13 has been tentatively interpreted as an ongoing planetary TDE \citep{Guo+2025}. These observations confirm that catastrophic star-planet interactions occur, but they also highlight a critical challenge: distinguishing between a direct merger and a tidal disruption. This distinction is not always straightforward, since the two channels—while physically distinct—can naturally occur in sequence when orbital decay is rapid, as the case of ZTF~SLRN-2020 might suggest.

In this work, we focus on modeling the specific observational signatures of planetary TDEs. We present 2D hydrodynamic simulations to model the post-disruption evolution of a Jupiter-like planet and a Neptune-like planet around a solar-mass star, and we predict the resulting observational signatures. We systematically investigate how the light curve morphology depends on both the progenitor's mass and its initial orbital eccentricity. In real astrophysical systems, these eccentricities can vary widely: slow tidal decay may produce nearly circular orbits, while planet--planet scattering, Kozai--Lidov cycles, or the injection of free-floating planets can generate highly eccentric or even near-parabolic trajectories (e.g., \citealp{Chatterjee+2008, Perets+2012, Kozai-Lidov2016}). These high-$e$ pathways are especially relevant, as they can bring a planet inside the Roche limit before any effective circularisation takes place. The goal is to provide a detailed theoretical framework to aid in the identification and classification of these events in current and future time-domain surveys, such as the Vera C. Rubin Observatory's Legacy Survey of Space and Time (LSST) \citep{Ivezi2019LSST}, and to establish clear observational criteria to distinguish them from direct star-planet mergers. Furthermore, automated tools like the ALeRCE expansion classifier are already incorporating TDEs into their taxonomy \citep{ALERCE+2025}, making theoretical templates crucial for distinguishing these events from direct star-planet mergers.

The paper is organised as follows. In Sect.~\ref{sec:Theoretical_framework}, we describe the theoretical framework and the setup of our hydrodynamic simulations. Sect.~\ref{sec:results} presents the main results, including the light curves, thermal evolution, and spectral signatures. In Sect.~\ref{sec:discussion}, we discuss the implications for observations and for distinguishing these events from other transients, and in Sect.~\ref{sec:conclusions} we summarise our conclusions.

\section{Theoretical framework}
\label{sec:Theoretical_framework}

\subsection{Tidal disruption or direct impact}
\label{sec:Tidal_disruption}
To establish the initial physical regimes for planet–star interactions, we first calculate the classical Roche limit, $\dR$. This planet-star separation distance sets the condition that determines whether a planet orbiting its host star undergoes a tidal disruption, in which the planet is stripped and its internal binding energy is released into a debris stream, or whether it instead collides directly with the stellar surface. If the Roche limit lies outside the stellar radius ($\dR > \Rs$), tidal forces tear the planet apart as it crosses this boundary, feeding a flow of debris that releases the gravitational energy which powers the observable transient. 

If $\dR > \Rs$ the planet reaches the stellar photosphere before disruption can occur, producing a very different outcome with no debris phase. It is important to note that the disruption of a planet in a circular orbit may proceed as a stable Roche lobe overflow (RLOF) rather than a catastrophic TDE, depending on the mass ratio and the planet's response to mass loss \citep{Lai+2010, Valsecchi+2015, Jackson+2017, Jia+2017}. However, as the orbit shrinks due to tidal dissipation, the system can eventually reach a terminal stage where mass transfer becomes unstable or the planet is rapidly engulfed \citep{Dosopoulou+2017}. In this work, our circular ($e=0$) model is intended to represent the hydrodynamical evolution of the debris once this catastrophic mass-loss regime is triggered. The pathway followed by the planet depends mainly on the ratio between the stellar and planetary densities, as given by the standard expression:

\begin{equation}
\dR =  k  \Rs \left( \frac{\rhos}{\rhop} \right)^{1/3},
\label{eq:roche}
\end{equation}
where $\Rs$ and $\rhos$ are the radius and mean density of the host star, and $\rhop$ is the mean density of the planet. The dimensionless coefficient $k$ depends on the internal rigidity of the satellite: $k = 2.44$ for fluid bodies (gas giants, super-puffs) and $k = 1.26$ for rigid bodies (terrestrial planets); see for instance \citealp{MurrayBook1999}.

We note, however, that these classical limits assume a homogeneous internal density. Realistic astrophysical bodies possess density gradients (e.g., a dense core surrounded by a tenuous envelope), implying that tidal disruption is not a sharp threshold but rather a process that can occur partially across a range of distances. This distinction is particularly relevant for the disruption of differentiated bodies, such as the Neptune-like model discussed later in Sect. \ref{sec:dichotomous effect}.

We model planetary archetypes orbiting a Solar-mass star ($\Rs = 0.00465\,\AU$, $\rhos = 1.41\,\mathrm{g\,cm^{-3}}$): Jupiter- and Saturn-like gas giants and low-density super-puffs. For this stellar density, the Roche-limit criterion summarised in Table~\ref{tab:roche_limits} places these planets in the tidal-disruption regime, whereas dense rocky planets would enter the star before being disrupted. This classification, however, is specific to Solar-type stars; the outcome can change considerably with the stellar density (see Fig.~1 in \citealt{Sucerquia+Montesinos2025}). 

\begin{table*}[]
\centering
\caption{Roche Limits and Interaction Status for Planetary Archetypes Orbiting a solar-mass Star.}
\label{tab:roche_limits}
\begin{tabular}{l c c c c c}
\hline\hline
Planet & Type & Density & Roche limit & Period at $\dR$ & Status \\
 & & [g/cm$^3$] & [AU] & [yr] & \\
\hline
Jupiter & fluid & 1.33 & 0.0116 & 0.0012 & Tidal Disruption \\
Neptune & fluid & 1.64 & 0.0108 & 0.0011 & Tidal Disruption \\
Saturn & fluid & 0.69 & 0.0144 & 0.0017 & Tidal Disruption \\
Super-puff & fluid & 0.05 & 0.0345 & 0.0064 & Tidal Disruption \\
Earth & rigid body & 5.51 & 0.0037 & 0.0002 & Direct Impact \\
\hline
\end{tabular}
\end{table*}

\subsection{TDE modeling}
\label{sec:hydro_sim}

To model the TDE, we performed 2D hydrodynamic simulations using the \texttt{FARGO3D} code \citep{Benitez-Llambay+2016}. We initialize the system with a solar-mass star and a Jupiter-like planet placed at its Roche limit. The simulation employs a thin, viscous $\alpha$-disk model  \citep{Shakura+Sunyaev1973} with a viscosity parameter $\alpha = 10^{-3}$ and an initial aspect ratio of $H/r = 0.05$.

Our model solves the time-dependent energy equation for an ideal gas with fixed adiabatic index $\gamma$ (=5/3), including viscous heating and radiative cooling, which together set the midplane  and the effective temperature via the local effective optical depth (details in subsection \ref{subsec:hydro}). The initial setup represents the moment the planet begins to be tidally disrupted, with its material being fed into the circumstellar environment. We use two temporal resolutions to capture the different evolutionary phases:
\begin{itemize}
    \item \textbf{Short-term evolution:} A high temporal resolution run covering the first $\sim$ 14 days, with outputs saved every  $\Delta t \sim 10^{-4} \rm{yr} \sim 1$ hour.  This sampling resolves the innermost (Jupiter) orbital period at $d_R$ ($\approx 10.5$ hours, see Table \ref{tab:roche_limits}) with approximately 10 snapshots per orbit, crucial for capturing the initial circularization dynamics. Such timescale is significantly faster than the typical cadence of all-sky surveys like LSST, but is well within the capabilities of targeted instruments like TESS \citep{TESS2015} or PLATO \citep{PLATO2014}.
    
    \item \textbf{Long-term evolution:} A lower temporal resolution run covering $\sim$1800 days (5 years), with outputs every $\sim 10^{-2}$ yr $\sim$ 3.65 days, to track the viscous decay phase. This cadence is well-suited to monitoring the viscous decay phase, which occurs over several years, and is compatible with the long-term monitoring strategies of surveys like LSST.
\end{itemize}

\subsection{Computational domain and initial conditions}

The computational grid spans a radial range of $r \in [1, 15]~R_\odot$ and an azimuthal range of $\phi \in [0, 2\pi]$, with a resolution of $N_r \times N_\phi = 384 \times 512$ cells. We confirmed the numerical convergence of our results by performing a higher-resolution test run with $N_{r} \times N_{\phi} = 720 \times 1024$ cells. We verified that the spatial morphology and temporal evolution of the primary hydrodynamic variables---surface density, velocity, and internal energy---remained consistent with the fiducial model. Consequently, the derived global observables, such as the bolometric luminosity and characteristic timescales, are robust and not artifacts of the grid resolution. Open boundary conditions were used at both the inner and outer radial boundaries, allowing material to flow freely off the computational grid without reflection.

The simulation is initialized by modeling the disk's surface density, $\Sigma$, as a localized enhancement, $\Delta\Sigma_{\rm clump}$, superimposed on a tenuous, power-law background medium:

\begin{equation}
    \Sigma(r,\phi) = \Sigma_0 \left( \frac{r}{R_0} \right)^{-p} + \Delta\Sigma_{\rm clump}.
    \label{eq:sigma_total}
\end{equation}

This background component represents a physical pre-existing circumstellar disk (e.g., the inner region of a protoplanetary disk). We set a power-law index $p=-1$ normalized to $\Sigma_0 = 100$ g cm$^{-2}$ at $R_0=1$ au. Integrating this profile over the computational domain yields a total background mass of $\approx 1 M_J$, which is comparable to the mass of the disruption progenitor. This setup is consistent with a scenario where the planet has arrived at the Roche limit via disk-driven migration.

The clump $\Delta\Sigma_{\rm clump}$ is centered at the respective Roche limit for each planetary model ($r_p \approx 0.012$~au for Jupiter and $r_p \approx 0.011$~au for Neptune, see Table~\ref{tab:roche_limits}), and its material is confined to the planet's Hill radius $R_H$, defined as:

\begin{equation}
    R_H = r_p \left( \frac{M_p}{3 M_\star} \right)^{1/3}.
    \label{eq:hill_radius}
\end{equation}

Within this region, the surface density enhancement follows:

\begin{equation}
\Delta\Sigma_{\rm clump}(r,\phi) = A \exp\left( -\frac{d^2}{2\sigma^2} \right) \quad \text{for} \quad d \le R_H,
\end{equation}
where $d^2 = (r - r_p)^2 + r^2(\phi - \phi_p)^2$ is the metric-consistent in-plane distance. The Gaussian width is set to $\sigma = R_H / 3$, and the amplitude $A$ is normalized to ensure the clump contains the total planetary mass, $M_p$.

Regarding the thermal initialization, the planetary clump is set with the same midplane temperature profile as the background disk at that location. Since the planet density is orders of magnitude higher than the background, this configuration is not in pressure equilibrium; rather, it represents an over-pressured region (see Eq. \ref{eq:Pressure}). This physically captures the state of the material immediately after the loss of the planet's self-gravity, driving the initial rapid expansion. Furthermore, due to the high density of the clump, the optical depth is large, resulting in a low initial effective temperature ($T_{eff} \propto T_{mid}/\tau_{eff}^{1/4}$), consistent with the ``cold'' appearance of the debris in the initial snapshots (see Fig. \ref{fig:disk_evolution}).

The initial velocity of the background gas disk is set to represent stable, near-circular rotation. The azimuthal velocity, $v_{\phi}$, corresponds to the Keplerian orbital speed, with a slight sub-Keplerian correction to account for the radial pressure gradient, while the radial velocity is set to zero apart from a small random noise component.

This setup represents the system at the onset of the tidal disruption, capturing the moment the planetary material is no longer self-gravitating and begins its hydrodynamic evolution. To explore the effects of the planet's initial orbit, we initialize the Gaussian clump representing the planetary material in two distinct configurations:

\textbf{Fiducial Model} (Circular Orbit, $e=0$): Our fiducial model represents a planet that has undergone slow migration, beginning its disruption from a quasi-circular orbit at the Roche limit. In this setup, the Gaussian clump is given the same initial velocity as the background gas disk at its location: a near-circular Keplerian speed with a slight correction for the pressure gradient. This models the moment the planetary gas loses its self-gravity and begins to spread and circularize from an initially compact state.

\textbf{Eccentric Model} ($e=0.5$): For comparison, we run a second model to study a planet disrupted from a plunging trajectory. In this case, the clump is initialized at the pericenter of an orbit with an eccentricity of $e=0.5$. The corresponding semi-major axis is given by $a = r_p/(1-e)$. At this point, the clump's radial velocity is zero ($v_r=0$), and its azimuthal velocity is set by an elliptical orbit at pericenter:

\begin{equation}
    v_{\phi} = \sqrt{\frac{G M_*}{a} \frac{1+e}{1-e}}.
\end{equation}

To ensure numerical stability at the start of the simulation, this eccentric velocity is not applied uniformly. Instead, it is gradually tapered within the Hill radius using a smoothing function that smoothly connects the clump's pericenter velocity at its center to the background disk's velocity at its edge.

These setups represent the physical state of the planetary material at the moment of its closest approach to the star. They provide the initial conditions for the tidal disruption event, which is followed first by the debris circularization process and later by the viscous accretion phase.

Comparing the evolution from these two distinct initial conditions allows us to isolate the effects of the initial orbital energy and eccentricity on the subsequent debris circularization process, the viscous accretion phase, and the resulting observational signatures.

\subsection{Thermodynamics}\label{subsec:hydro}

We solve the non-stationary equation for the gas energy density, $\epsilon$, given by:
\begin{equation}
\frac{\partial \epsilon}{\partial t} + \nabla \cdot (\epsilon \mathbf{v}) = -P \nabla \cdot \mathbf{v} + Q^{+}_{v} - Q^{-},
\end{equation}
where $\mathbf{v}$ is the gas velocity, $P$ the gas pressure, $Q^{+}_{v}$ the viscous heating rate per unit area, and $Q^{-}$ the radiative cooling of the disk. 

We compute the full viscous heating term following \cite{D'Angelo-2003}, and adopt the $\alpha$-viscosity prescription of \cite{Shakura+Sunyaev1973} to model the kinematic turbulent viscosity.

Radiative cooling is implemented using the effective optical depth approximation:
\begin{equation}
Q^{-} = \frac{2 \sigma_{\rm SB} T_{\rm mid}^4}{\tau_{\rm eff}},
\label{eq:cooling}
\end{equation}
where $\sigma_{\rm SB}$ is the Stefan-Boltzmann constant and $T_{\rm mid}$ the midplane temperature. The effective optical depth $\tau_{\rm eff}$, is computed with the grey LTE formula of \citet{Hubeny1990}:

\begin{equation}\label{eq:eff_opacity}
\tau_{\rm eff} = \frac{3}{8} \tau_R  +  \frac{\sqrt{3}}{4}  + \frac{1}{4 \tau_P},
\end{equation}
with \(\tau_R \equiv \tfrac{1}{2}(\kappa_{\rm abs}+\kappa_{\rm es})\Sigma\) and \(\tau_P \equiv \tfrac{1}{2}\kappa_{\rm abs}\Sigma\). Here, \(\kappa_{\rm es}\) is the electron scattering opacity, given by \(\kappa_{\rm es}=0.2(1+X)\) cm\(^2\) g\(^{-1}\) for a hydrogen mass fraction \(X\) (we adopt $X = 0.7$).

For the absorption opacity $\kappa_{\rm abs}$, we adopt a smoothed implementation of the Rosseland-mean opacity following \citet{Bell-Lin1994} in the dust and H$^{-}$/molecular regimes, and we attach at high temperatures a Kramers bound-free/free-free continuation ($\kappa \propto \rho\,T_{\rm mid}^{-7/2}$) matched by continuity at a transition temperature $T_k$. This choice is physically motivated by the presence of the pre-existing circumstellar material and the high metallicity of the planetary debris, ensuring that the thermal evolution reflects the microphysics of a relatively dense protoplanetary disk ($M_{\rm disk} \sim 1 M_J$) rather than a tenuous plasma.

To close the system of equations, we assume an ideal equation of state:

\begin{equation}\label{eq:Pressure}
P = \Sigma T_{\rm mid} R,
\end{equation}
with $T_{\rm mid}$ the gas mid temperature and $R = k_{\mathrm{B}}/(\mu m_{p})$, where $k_{\mathrm{B}}$ is the Boltzmann constant, $\mu$ the mean molecular weight of the gas (we adopt $\mu = 2.35$), and $m_{p}$
the proton mass. The relation between the thermal energy density and the temperature is then:

\begin{equation}
\epsilon = \frac{\Sigma T_{\rm mid} R}{\gamma - 1},
\end{equation}
where $\gamma$ is the adiabatic index.

We note that our assumption of a constant mean molecular weight $\mu=2.35$ strictly describes the initial cold, neutral state of the planetary debris and background gas. If the gas is shock-heated to temperatures $T \gg 10^4$ K, hydrogen ionization would physically lower $\mu$ to $\sim 0.6$. While keeping $\mu$ constant implies an underestimation of the gas pressure and pressure scale-height in such high-temperature regimes (since $P \propto \mu^{-1}$ and $H \propto \mu^{-1/2}$), the radiative cooling in our model is controlled by a realistic variable opacity (Eq. \ref{eq:eff_opacity}) that correctly captures the transitions between molecular, atomic, and ionized regimes. Since the dominant high-temperature opacity ($\kappa_{es}$) is independent of density, the cooling rates remain physically plausible even if the pressure scale height is underestimated.


\subsection{Multi-band light curves evolution}
\label{sec:post_processing}

The energy equation provides the gas midplane temperature, $T(r,\phi,t)$, from which we compute the local effective temperature, $T_{\rm eff}$. This quantity represents the temperature of the disk's emitting surface and is calculated using the effective opacity (eq. \ref{eq:eff_opacity}):

\begin{equation}\label{eq:Teff_relation}
T_{\rm eff}^4(r,\phi,t) = T_{\rm mid}^4(r,\phi,t) / \tau_{\rm eff}(r,\phi,t).
\end{equation}

This effective temperature is then used as the input for a blackbody specific intensity model to compute the observable flux from each surface element. For a photometric band \(b\) with central wavelength \(\lambda_b\), the Planck function is
\beq{eq:planck_lambda}
B_{\lambda_b}\!\left(T\right) \;=\; \frac{2\,h\,c^2}{\lambda_b^{5}}\;\left[\exp\!\left(\frac{h c}{\lambda_b k_{\!B} T_{\rm eff}}\right)-1\right]^{-1}.
\eeq

Assuming Lambertian emission from each surface element, the monochromatic luminosity is the surface integral

\beq{eq:Llambda_cont}
L_{\lambda_b}(t) \;=\; 2 \int_{A} \pi\, B_{\lambda_b}\!\left(T_{\rm eff}(r,\phi,t)\right)\; \mathrm{d}A,
\eeq
with midplane area element in polar coordinates ${d}A \;=\; r\,\mathrm{d}r\,\mathrm{d}\phi$. We include emission from both disk faces consistently with the cooling prescription (Eq.~\ref{eq:cooling}), hence the factor 2.

To characterize the global thermal state of the non-isothermal disk, we define two distinct average effective temperatures. The first is the luminosity-weighted average effective temperature, $\langle T_{\rm eff} \rangle_L$, which serves as an analogue to the colour temperature by giving greater weight to the hotspots that dominate the total flux:
\begin{equation}\label{eq:Tmean_Luminosity}
\langle T_{\rm eff} \rangle_L = \frac{\int_A T_{\rm eff} \cdot (\sigma T_{\rm eff}^4) \, dA}{\int_A \sigma T_{\rm eff}^4 \, dA} = \frac{\int_A T_{\rm eff}^5 \, dA}{\int_A T_{\rm eff}^4 \, dA},
\end{equation}
where $T_{\rm eff}$ is the local effective temperature (eq. \ref{eq:Teff_relation}) at each surface element $dA$.

The second metric is the area-weighted average effective temperature, $\langle T_{\rm eff} \rangle_A$, which represents the fourth root of the area-weighted mean of $T_{\rm eff}^4$:
\begin{equation}\label{eq:Tmean_Area}
   \langle T_{\rm eff} \rangle_A = \left( \frac{\int_A T_{\rm eff}^4 \, dA}{\int_A dA} \right)^{1/4}.
\end{equation}

We use these quantities alongside the total bolometric luminosity to analyze the system's evolutionary track in the L-T plane. Furthermore, the comparison between these two average temperatures provides a powerful diagnostic for the dominant heating mechanism in our different simulation models, as will be shown in Section~\ref{sec:results}.

At distance \(d\), the observed monochromatic flux is
\beq{eq:Flambda}
F_{\lambda_b}(t) \;=\; \frac{L_{\lambda_b}(t)}{4\pi d^2},
\eeq
with $d = 160$ pc adopted for the source distance.

Colour indices are computed from band flux ratios. For bands \(b_1\) and \(b_2\),
\beq{eq:color}
(b_1 - b_2)(t) \;=\; -2.5\log_{10}\!\left(\frac{F_{b_1}(t)}{F_{b_2}(t)}\right) \;+\; Z_{b_1 b_2},
\eeq
where \(Z_{b_1 b_2}\) is the adopted zeropoint constant. 

For our analysis, we adopt the standard Johnson-Cousins ($V, R$) and Johnson ($J, K, L, M$) photometric systems. Specifically, we compute the flux at the central wavelengths of $V$ ($0.55\,\mu$m), $R$ ($0.64\,\mu$m), $J$ ($1.25\,\mu$m), $K$ ($2.20\,\mu$m), $L$ ($3.50\,\mu$m), and $M$ ($4.80\,\mu$m). Note that we use the standard $K$-band center rather than $K_s$, and the standard $L$-band rather than $L^\prime$.

To generate synthetic images (see Sect.~\ref{sec:morphology}), we post-process the hydrodynamic outputs using the radiative transfer code \texttt{RADMC-3D} \citep{radmc3d}. Given that the peak disk temperatures ($T > 10^4$ K, as confirmed a posteriori) far exceed the sublimation threshold for silicate grains ($\sim 1500$ K), the primordial planetary dust is expected to be rapidly vaporized. Consequently, we adopt a highly depleted dust-to-gas ratio of $10^{-8}$, assuming standard olivine grains with sizes ranging from $0.1$ to $10\,\mu$m for this trace component. This choice ensures that the resulting synthetic maps trace the optically thin structure of the hot debris using the hydrodynamic temperature field, thus employing \texttt{RADMC-3D} strictly as a post-processing imaging tool consistent with the gas-dominated cooling prescription (Eq.~\ref{eq:eff_opacity}) used during the simulation.

\subsection{Temporal evolution timescales}
\label{sec:timescales}

The evolution of the disrupted planetary material occurs over two distinct timescales.
The initial, rapid \textit{dynamical timescale}, $t_{\rm dyn}$, governs the catastrophic disruption of the planet and the formation of a transient disk. Once the planet overflows its Roche lobe, it can be engulfed by the star, leading either to stable mass transfer or to the planet’s rapid disruption into an accretion disk, giving rise to optical, UV and X-ray transients from planet–star mergers \citep{Metzger+2012}. This phase is characterized by the redistribution of material under the star's gravity, producing a quasi-Keplerian disk within hours to days.

Following this, the long-term evolution is governed by the much slower \textit{viscous timescale}, $t_{\rm visc}$. Viscous torques within the disk transport angular momentum outward, allowing material to lose energy and accrete onto the star. This process drives the circularization of the debris and powers the observable transient light curve over weeks, months or years. Our simulations are designed to capture both the initial dynamical phase and the subsequent viscous evolution.

\section{Results}
\label{sec:results}

\begin{figure}[]
    \centering
    \includegraphics[width=\linewidth]{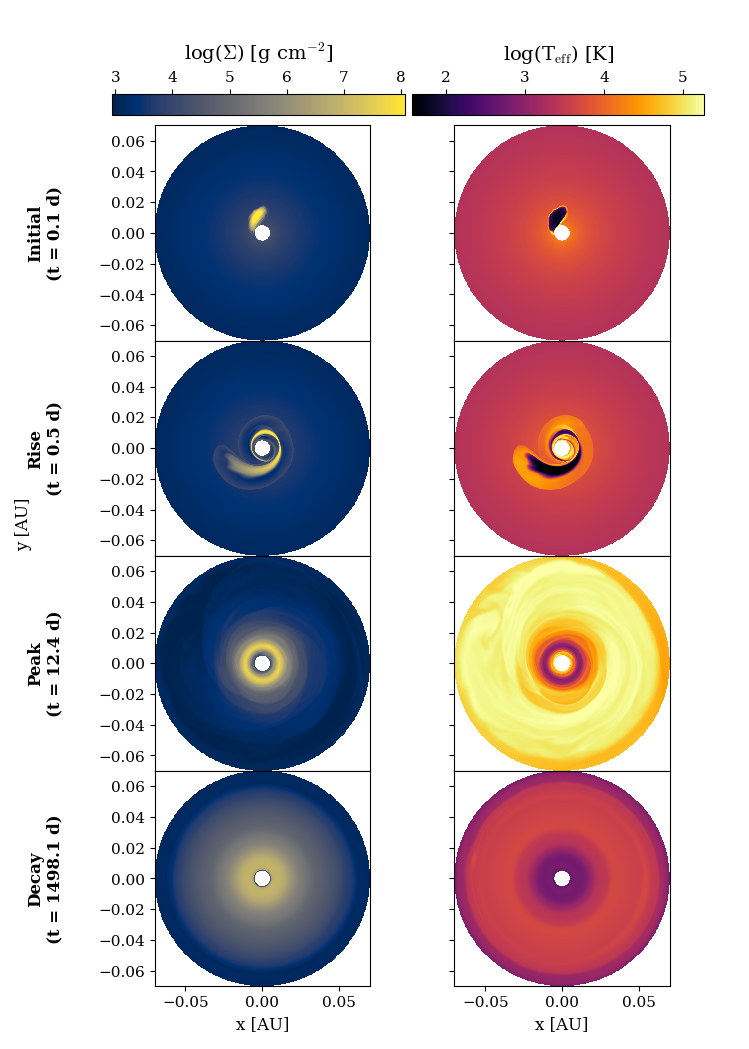}
    \caption{Evolution of the surface density (left column) and effective temperature (right column) for the fiducial Jupiter model ($e=0$). Each row corresponds to a key epoch: the initial compact clump, the shock-heating phase during the rising TDE, the hot peak disk phase, and the viscously relaxed state at late times.}
    \label{fig:disk_evolution}
\end{figure}

\subsection{Surface density and thermal evolution of the disk}

The evolution of the disk's physical structure provides direct insight into the dissipation of the planetary debris. Figure~\ref{fig:disk_evolution} illustrates the spatial distribution of surface density ($\Sigma$) and effective temperature ($T_{\rm mid}$) for the fiducial Jupiter model at four key evolutionary epochs.

\begin{itemize}
    \item \textbf{Initial (t=0.1 d):} Immediately after the disruption, the planetary material is concentrated in a dense, compact clump with a well-defined density peak; the surrounding material is at an extremely low density. This initial clump is characterized by its low entropy: it has a significantly higher density, but its internal temperature ($T_{\rm effective} \lesssim 10^2$ K) is notably cooler than the tenuous, pre-existing circumstellar disk. This is visible in the temperature map as a distinct cold spot.

    \item \textbf{Rise (t=0.5 d):} The clump is rapidly sheared by differential rotation, spreading into a crescent-like structure. The material's density remains confined within this compact feature, and its dense interior remains cool, preserving its initial low entropy. However, its leading edge moves at supersonic speeds, creating a strong bow shock as it impacts the surrounding gas. This shock front is visible as an intensely heated arc in the temperature map, where kinetic energy is violently dissipated reaching temperatures about $\sim 5 \times 10^4$ K, hotter than the pre-existing low mass circumstellar disk.

    \item \textbf{Peak (t=12.4 d):} Within 12.4 days, the planetary material has spread significantly. The circularization process has distributed the density, which is no longer in a compact clump but has formed an extended, luminous, and nearly axisymmetric accretion disk. The entire structure is in a hydrodynamically turbulent state. The effective temperature map reveals that this material expansion has heated a broad area, with high effective temperatures ($\sim 3 \times 10^5$ K) now extending throughout the newly formed disk.

    \item \textbf{Decay (t=1497.3 d):} After extended viscous evolution, the system has relaxed. The density has become nearly axisymmetric and has spread to larger radii, resulting in a much lower average surface density than at the peak. The disk has cooled significantly, with the effective temperature dropping by approximately two orders of magnitude. A clear radial temperature gradient is established, consistent with a viscously heated accretion disk in a quasi-steady state.
\end{itemize}

This morphological sequence shows the fundamental transition of the system: from an initial, low-entropy clump to a hot, shock-dominated turbulent disk, and finally to a cooler, viscously evolving structure.

\subsection{Morphological evolution of the infrared emission}
\label{sec:morphology}

\begin{figure*}[!h]
   \centering
   \includegraphics[width=0.8\linewidth]{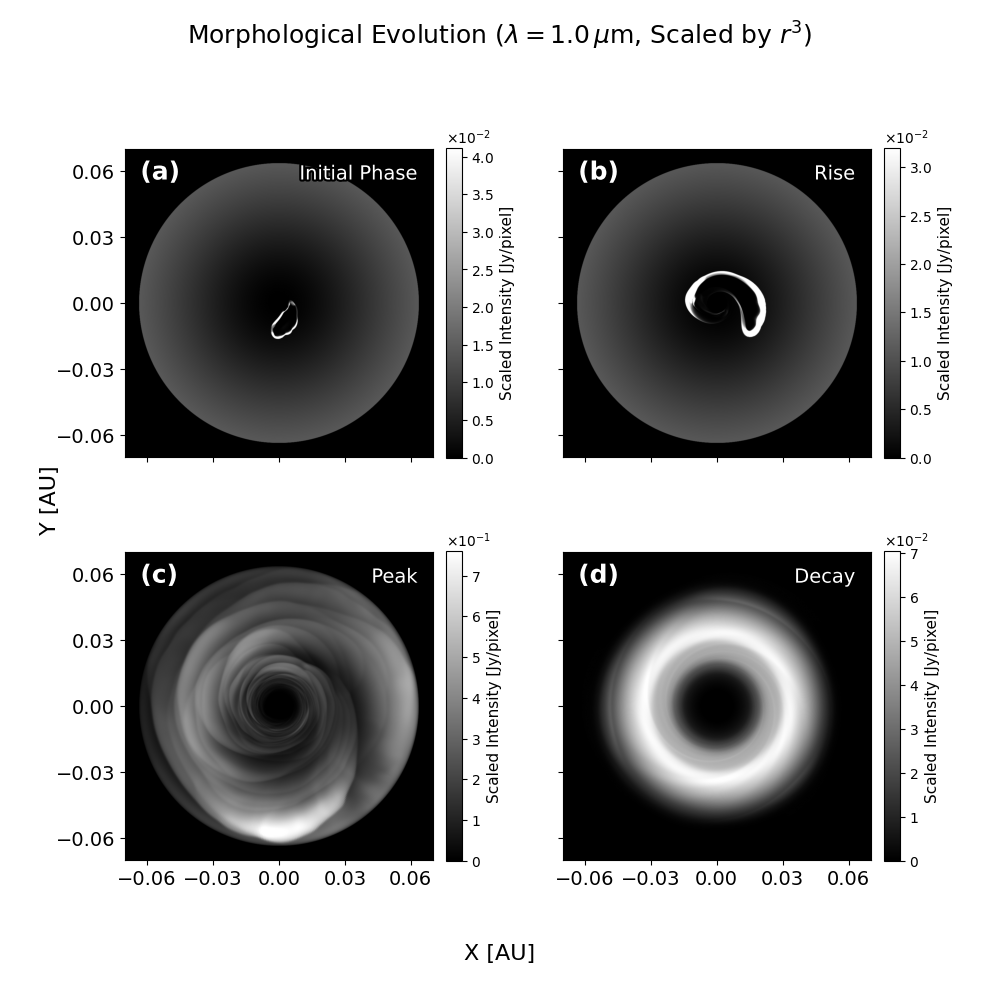} 
   \caption{Synthetic imaging of the fiducial Jupiter TDE at $\lambda = 1.0\,\mu$m observed from $d=140$~pc. The four panels show the morphological evolution from the initial compact debris (a) and the formation of the primary spiral arm during the rise (b), to the turbulent peak (c) and the relaxed disk (d). To enhance the visibility of the emission structure, the intensity is scaled radially by $(r/0.07\,\mathrm{au})^3$. For reference, the total physical flux density of the system at the peak epoch (c) is $\sim 1.7 \times 10^2$~Jy.}
   \label{fig:morphology}
\end{figure*}

While the surface density and temperature maps reveal the hydrodynamic state of the debris, the observational appearance of the system is determined by the emergent flux. Figure~\ref{fig:morphology} presents synthetic images of the fiducial Jupiter model at $\lambda = 1.0\,\mu$m (near-infrared $J$-band), generated using the radiative transfer methodology described in Sect.~\ref{sec:post_processing}. These maps trace the intensity of the thermal emission at a distance of $d=140$~pc.

To visualize the faint, extended spiral structures against the orders-of-magnitude brighter central accretion zone, the intensity maps in Fig.~\ref{fig:morphology} have been radially scaled by a factor of $(r/r_{\rm max})^3$. Without this scaling, the image would be dominated solely by a point-like central source. We emphasize, however, that the flux values discussed below refer to the physical, unscaled emission.

The morphological evolution follows four distinct phases:
\begin{itemize}
    \item \textbf{Initial Phase (Panel a, t=0.1 d):} The emission is faint ($F_{\rm tot} \approx 6.1$~Jy) and confined to the immediate vicinity of the disrupting planet. The material has not yet spread significantly, appearing as a compact, localized source.
    
    \item \textbf{Rise (Panel b, t=0.5 d):} As the debris stream begins to circularize, the emission traces the formation of the primary spiral arm. The shocked material at the leading edge of the stream becomes a significant source of radiation, with the total flux rising slightly to $\approx 6.4$~Jy.
    
    \item \textbf{Peak (Panel c, t=12.4 d):} This epoch corresponds to the maximum bolometric luminosity. The image reveals a fully developed, highly turbulent accretion disk. The spiral shocks have effectively redistributed the material, creating a structure that extends out to $\sim 0.07$~au. The total integrated flux at this wavelength reaches $\sim 1.7 \times 10^2$~Jy, driven by the high effective temperatures throughout the extended disk surface.
    
    \item \textbf{Decay (Panel d, t=1497.3 d):} In the late evolution, the turbulent spiral structures have subsided. The emission morphs into a smooth, axisymmetric ring-like structure. Although the disk remains extended, its cooling leads to a drop in the total infrared flux to $\sim 17.4$~Jy, consistent with the viscous fading of the system.
\end{itemize}

This sequence shows that the photometric peak of the light curve coincides with the phase of maximum structural complexity and spatial extent of the debris disk.

\subsection{Bolometric light curves}\label{subsec:Bolometric}

\begin{figure*}[]
    \centering
    \includegraphics[width=0.49\linewidth]{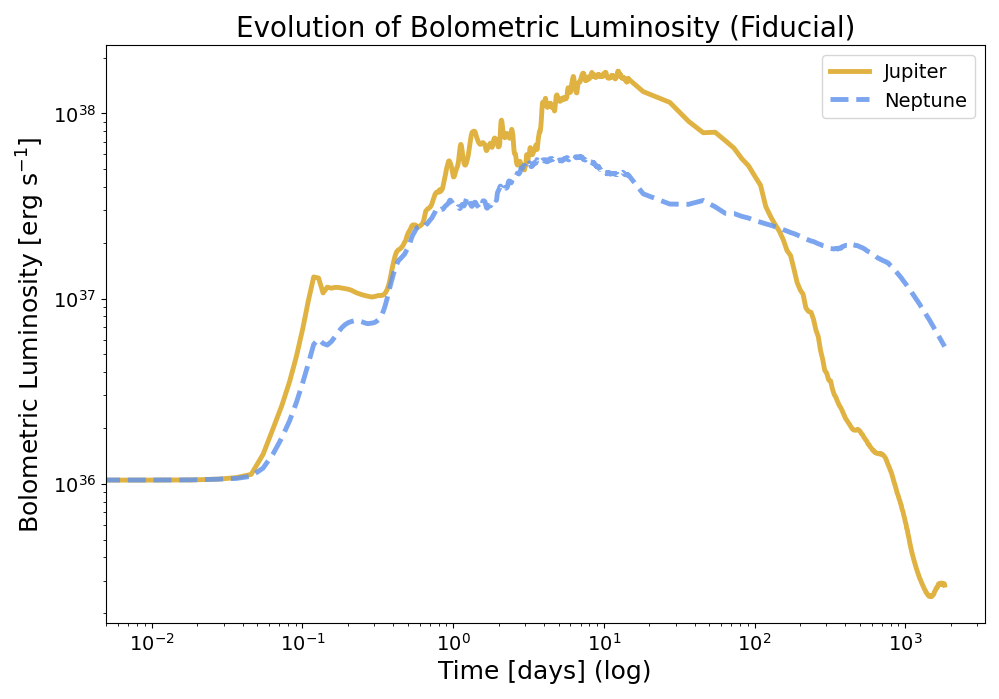}
    \includegraphics[width=0.49\linewidth]{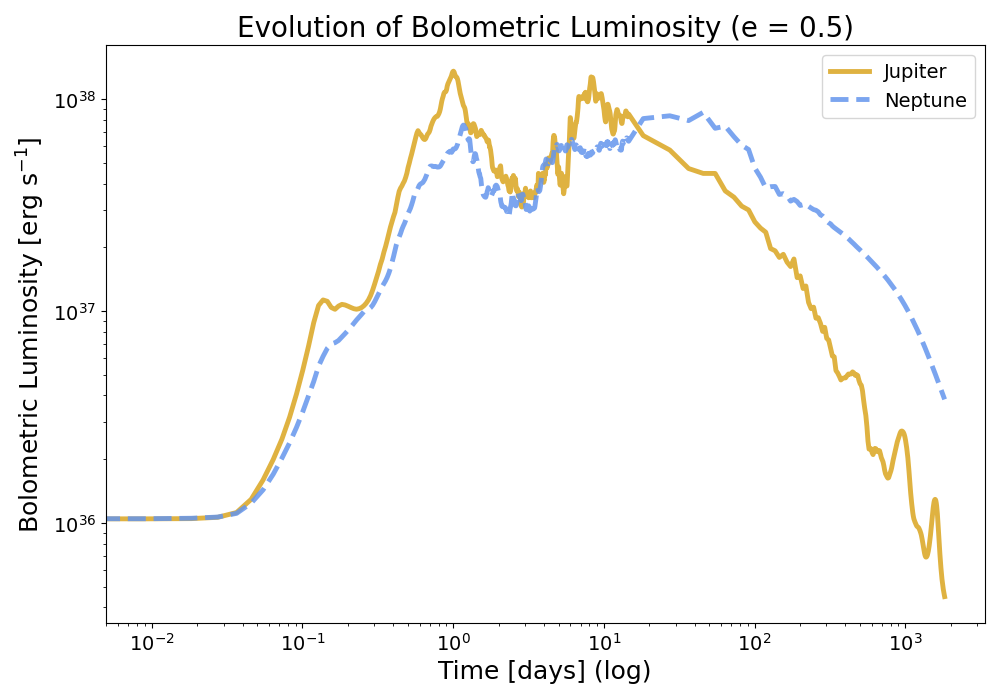}
    \caption{
    Comparison of bolometric light curves. 
    \textbf{Left:} Fiducial models ($e=0$). The Jupiter model (gold) shows a higher peak and faster decay, while the Neptune model (blue) is fainter but longer-lived.
    \textbf{Right:} Eccentric models ($e=0.5$). The effect of eccentricity is markedly different for Jupiter (faster, more volatile peak) and Neptune (slower, broader, and more luminous peak).
    }
    \label{fig:bolometric_comparison}
\end{figure*}

Figure~\ref{fig:bolometric_comparison} presents the bolometric light curves for all four simulated scenarios, comparing the fiducial ($e=0$, left panel) and eccentric ($e=0.5$, right panel) cases. The evolution reveals significant morphological differences driven by both the progenitor planet's mass and its initial orbital eccentricity.

For the fiducial ($e=0$) simulations, the Jupiter model (gold line, left panel) produces a bright transient that rises slowly, reaching its peak luminosity of $L_{\rm peak}=1.69\times10^{38}$~erg~s$^{-1}$ at $t_{\rm peak}\approx 12.4$~days. In contrast, the fiducial Neptune model (blue line, left panel) is fainter and peaks slightly earlier ($t_{\rm peak}\approx 7.0$~days), but exhibits a much more gradual decay phase.

Initial orbital eccentricity introduces dramatic and distinct changes (right panel). For Jupiter, a plunging orbit ($e=0.5$) results in a much faster and more violent event. The rise to the initial peak is an order of magnitude faster ($t_{\rm peak} \approx 1.0$~day), and the subsequent evolution is highly volatile, exhibiting a clear double-peaked structure—with a second, broader peak occurring around $\sim$10 days—a stark contrast to the smooth plateau of its circular-orbit counterpart.

Interestingly, eccentricity affects Neptune in the opposite manner. Rather than accelerating the event, the $e=0.5$ orbit \textit{delays} the peak to $t_{\rm peak} \approx 45.7$~days, resulting in a transient that is both brighter and more prolonged than its fiducial equivalent. This dichotomous effect highlights a complex interplay between the initial orbital energy and the planet's internal structure in shaping the observable characteristics of the disruption.

A comprehensive summary of the key quantitative properties for all four models is presented in Table~\ref{tab:models_comparison}. To characterize the light curve shapes, we fit the rise and decay phases to a power law, $L \propto t^{\alpha}$, to find the indices $\alpha_{\rm rise}$ and $\alpha_{\rm decay}$, respectively. These values quantify the visual differences seen in Figure~\ref{fig:bolometric_comparison}, such as the steep decay of the fiducial Jupiter model ($\alpha_{\rm decay} \approx -1.76$) versus the very shallow decay of its Neptune counterpart ($\alpha_{\rm decay} \approx -0.58$). The table also lists other key metrics, including the total radiated energy ($E_{\rm rad}$). Notably, the Neptune models consistently radiate more total energy than their Jupiter counterparts in both the circular ($2.26\times10^{45}$~erg vs. $1.14\times10^{45}$~erg) and eccentric ($2.66\times10^{45}$~erg vs. $9.32\times10^{44}$~erg) scenarios. This occurs because the total radiated energy ($E_{\rm rad} = \int L(t) dt$) is dominated by the long, shallow decay of the Neptune models, which (despite a lower peak) radiate for a longer duration than the brighter, shorter-lived Jupiter transients. Finally, the table includes the time for the flux to decay by factors of 10 and 100 ($t_{/10}$, $t_{/100}$).

Finally, we note that the Neptune models exhibit a higher overall radiative efficiency (ratio of radiated energy to available potential energy) compared to the Jupiter cases. This result, while initially counterintuitive given the lower peak luminosity, is attributable to the lower optical depth of the debris disk formed by the lower-mass planet. The lower density allows for more efficient radiative cooling, whereas the optically thicker Jupiter-like disk traps a larger fraction of the dissipated energy, which is then advected rather than radiated immediately.

\begin{table*}[h!]
\centering
\caption{Comparison of Intrinsic Scalar Properties for all Simulation Models. The parameters $t_{/10}$, $t_{/100}$, and $t_{/1000}$ represent the time in days, measured since the peak, for the bolometric luminosity to decay by a factor of 10, 100, and 1000, respectively.}
\label{tab:models_comparison}
\begin{tabular}{l c c c c}
\hline\hline
Quantity & Jupiter ($e=0$) & Neptune ($e=0$) & Jupiter ($e=0.5$) & Neptune ($e=0.5$) \\
\hline
$t_{\rm peak}$ [d] & 12.39 & 7.01 & 1.00 & 45.65 \\
$L_{\rm peak}$ [erg s$^{-1}$] & $1.69\times10^{38}$ & $5.84\times10^{37}$ & $1.36\times10^{38}$ & $8.67\times10^{37}$ \\
$t_{/10}$ (since peak) [d] & 170.2 & 1755.1 & 209.0 & 1113.8 \\
$t_{/100}$ (since peak) [d] & 517.1 & > 1800 & 1094.6 & > 1800 \\
$t_{/1000}$ (since peak) [d] & > 1800 & > 1800 & > 1800 & > 1800 \\
$E_{\rm rad}$ [erg] & $1.14\times10^{45}$ & $2.26\times10^{45}$ & $9.32\times10^{44}$ & $2.66\times10^{45}$ \\
$\alpha_{\rm rise}$ & 0.66 & 0.58 & 1.41 & 0.36 \\
$\alpha_{\rm decay}$ & -1.76 & -0.58 & -1.33 & -0.90 \\
$\langle T_{\rm eff}\rangle_L(t_{\rm peak})$ [K] & $4.84\times10^{4}$ & $4.38\times10^{4}$ & $8.54\times10^{4}$ & $4.07\times10^{4}$ \\
$\langle T_{\rm eff}\rangle_A(t_{\rm peak})$ [K] & $1.22\times10^{5}$ & $9.4\times 10^{4}$ & $1.16\times10^5$ & $1.03\times 10^5$ \\
\hline
\end{tabular}
\end{table*}

\subsection{Thermal evolution}

\begin{figure*}[]
    \centering
    \includegraphics[width=0.49\linewidth]{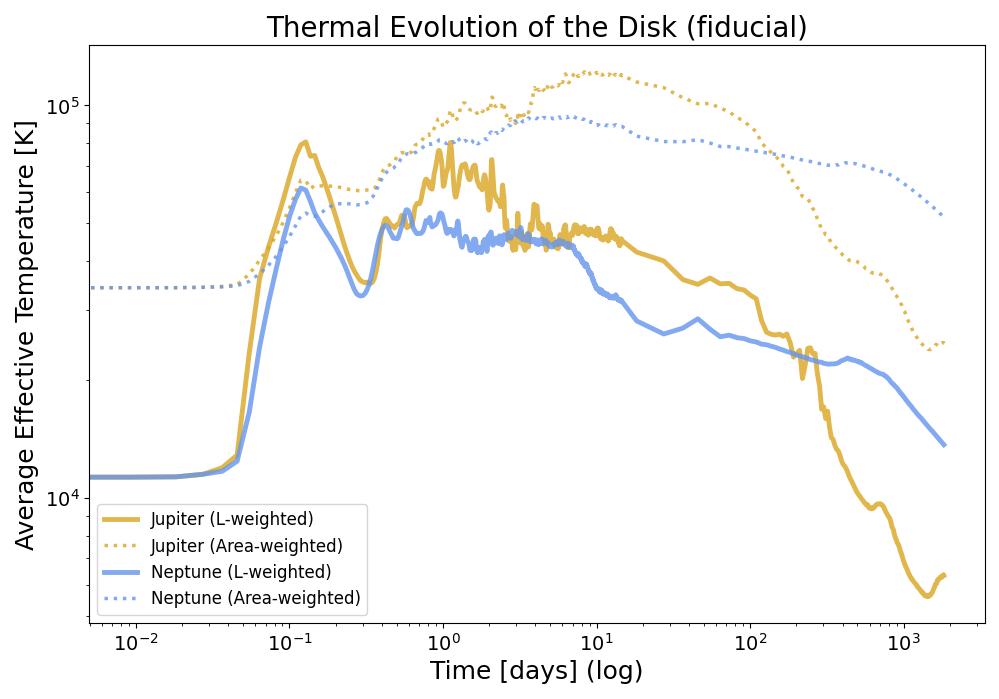}
    \includegraphics[width=0.49\linewidth]{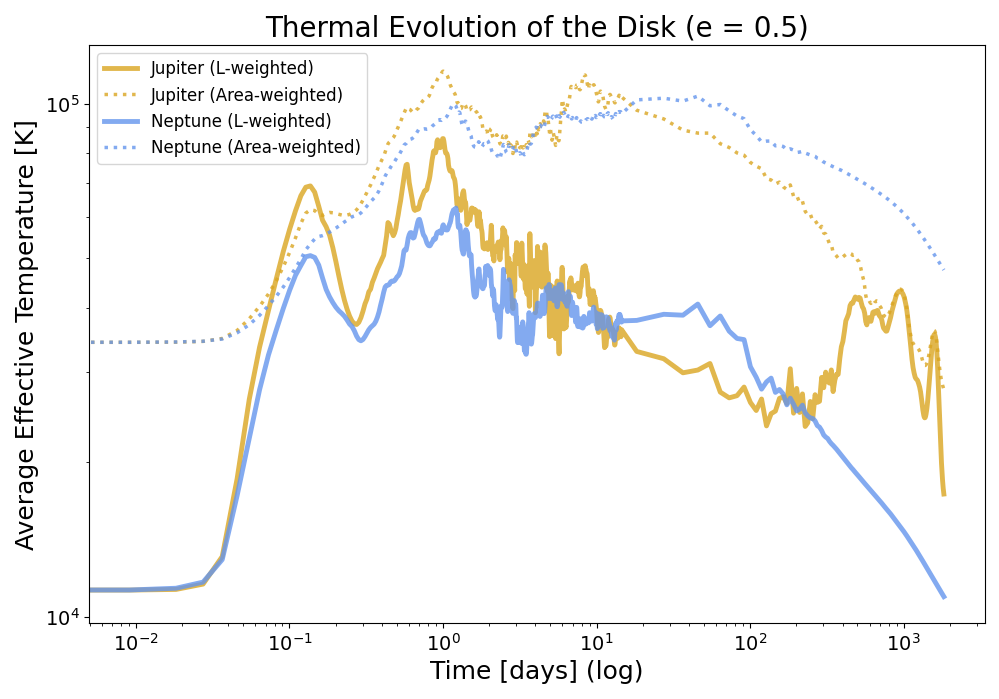}
    \caption{
    Comparison of the thermal evolution. 
    \textbf{Left:} In the fiducial ($e=0$) run, the temperature evolution is smooth post-peak. 
    \textbf{Right:} In the eccentric ($e=0.5$) run, the evolution is highly variable and spiky, especially for the Jupiter model, which reaches a higher peak temperature, indicating a more violent initial energy dissipation.
    }
    \label{fig:thermal_evolution_comparison}
\end{figure*}

\begin{figure*}[]
    \centering
    \includegraphics[width=0.49\linewidth]{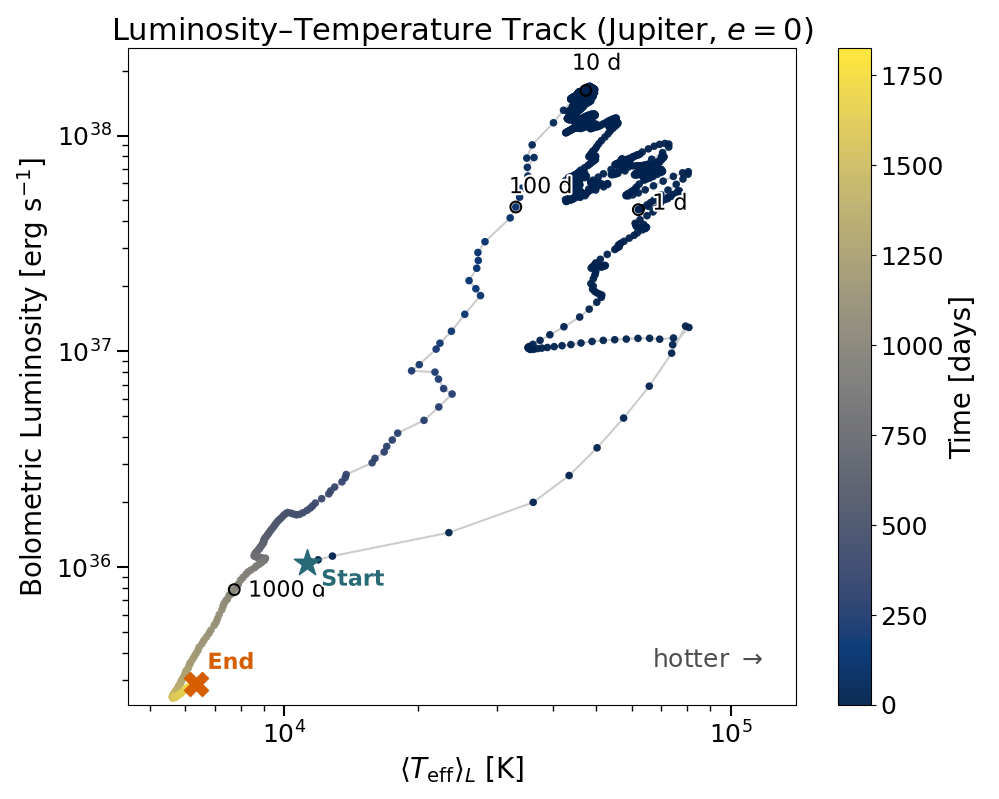}
    \includegraphics[width=0.49\linewidth]{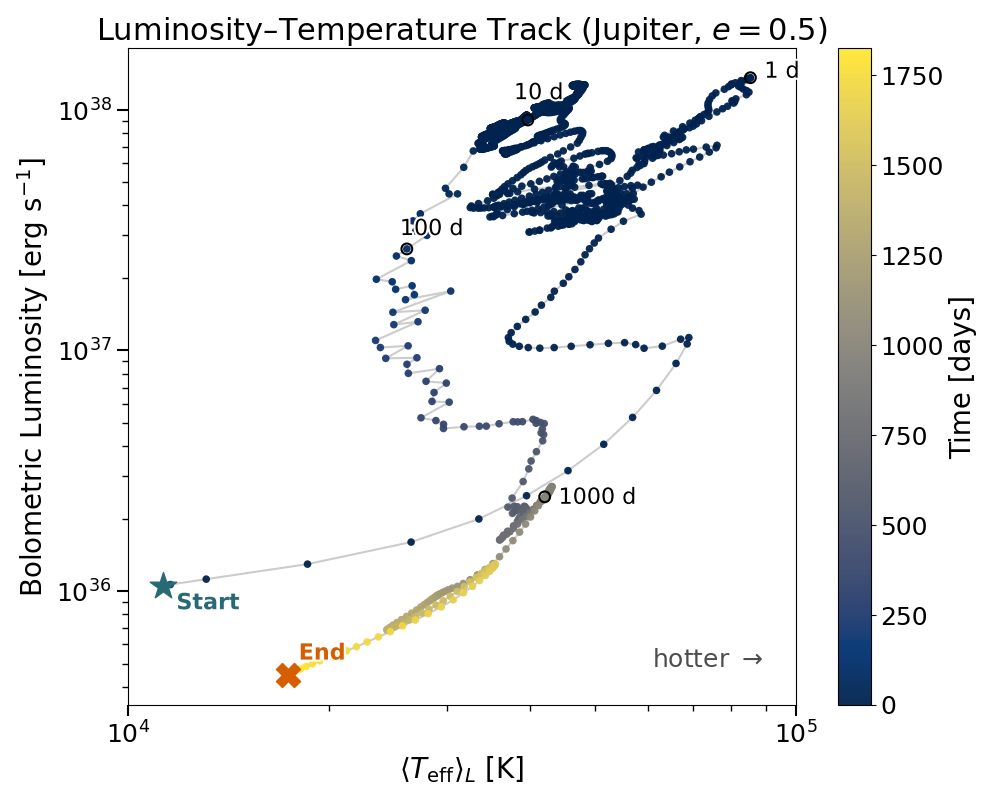}
    \caption{
    Luminosity--temperature tracks for the Jupiter model. 
    \textbf{Left:} The fiducial ($e=0$) case shows a relatively smooth track with well-defined loops near the peak, characteristic of a quasi-stable evolution. 
    \textbf{Right:} The eccentric ($e=0.5$) case displays a substantially more scattered and chaotic track, reflecting a highly unstable hydrodynamic evolution following the initial disruption.
    }
    \label{fig:L_T_comparison}
\end{figure*}

The thermal properties of the disk reveal the physical nature of the heating and cooling processes, which are strongly dependent on the initial conditions. We first examine the temporal evolution of the average effective temperatures in Figure~\ref{fig:thermal_evolution_comparison}, followed by the system's path in the luminosity-temperature (L-T) diagram in Figure~\ref{fig:L_T_comparison}.

Figure~\ref{fig:thermal_evolution_comparison} plots the evolution of the average effective temperatures for the fiducial ($e=0$, left panel) and eccentric ($e=0.5$, right panel) models. A key diagnostic shown here is the comparison between the luminosity-weighted average temperature, $\langle T_{\rm eff} \rangle_L$, and the area-weighted average, $\langle T_{\rm eff} \rangle_A$. In all four models, we find that $\langle T_{\rm eff} \rangle_A$ (dotted lines) significantly exceeds $\langle T_{\rm eff} \rangle_L$ (solid/dashed lines) around the luminosity peak. For the fiducial Jupiter model at its bolometric peak, $\langle T_{\rm eff} \rangle_L \approx 4.8\times10^4$~K, while the area-weighted temperature reaches a much higher value of $\langle T_{\rm eff} \rangle_A \approx 1.22\times10^5$~K. This discrepancy indicates that the emission is not dominated by a few small, bright hotspots, but rather by a widespread, global heating of the disk surface, a characteristic feature of the disruption process across all tested scenarios.

We can also interpret this discrepancy in terms of optical depth effects. As seen in the peak phase of the simulation, the regions of highest surface density (innermost radii) coincide with a suppression in the effective temperature. This occurs because the high optical depth ($\tau \propto \Sigma$) in these dense zones effectively traps the dissipated heat, forcing the emergent flux ($F \propto T_{\rm mid}^4/\tau$) to be lower than what the high internal temperatures would suggest. Consequently, the luminosity-weighted temperature is damped by these optically thick regions, falling below the area-weighted average which is dominated by the more extended, optically thinner, and brighter parts of the disk.

The fiducial ($e=0$) models exhibit a relatively smoother thermal evolution; after peaking, the temperature evolution is correspondingly gradual (left panel). In contrast, the eccentric models display a significantly more violent thermal evolution (right panel). The eccentric ($e=0.5$) Jupiter model, in particular, reaches a hotter luminosity-weighted peak temperature ($\langle T_{\rm eff}\rangle_L \approx 8.5\times10^{4}$~K) compared to its fiducial counterpart ($\approx 4.8\times10^{4}$~K). Its temperature evolution is highly chaotic and spiky, mirroring the volatile nature of its bolometric light curve.

This physical behaviour is reflected in the L-T diagrams shown in Figure~\ref{fig:L_T_comparison}. For the fiducial Jupiter model (left panel), the system follows a smoother evolutionary track after an initial complex phase, settling onto a well-defined cooling sequence. The eccentric Jupiter model (right panel), however, displays a substantially more scattered and chaotic track. This instability, which is especially prominent near the peak, confirms that the excess orbital energy associated with the plunging trajectory (specifically, its high kinetic energy at pericenter relative to the local circular speed) must be dissipated through a more rapid and hydrodynamically unstable process, leading to a hotter and more chaotic thermal state before the disk viscously relaxes.


\subsection{Intrinsic spectral and observed photometric signatures}

To provide a framework for observations, we now focus on the detailed spectral and photometric evolution of the fiducial Jupiter model, which serves as our primary case study. We first examine the intrinsic spectral energy distribution (SED) of the transient disk and then connect it to the resulting multi-band light curves and colour evolution.

\subsubsection{Spectral energy distribution}

\begin{figure}[h!]
    \centering
    \includegraphics[width=\linewidth]{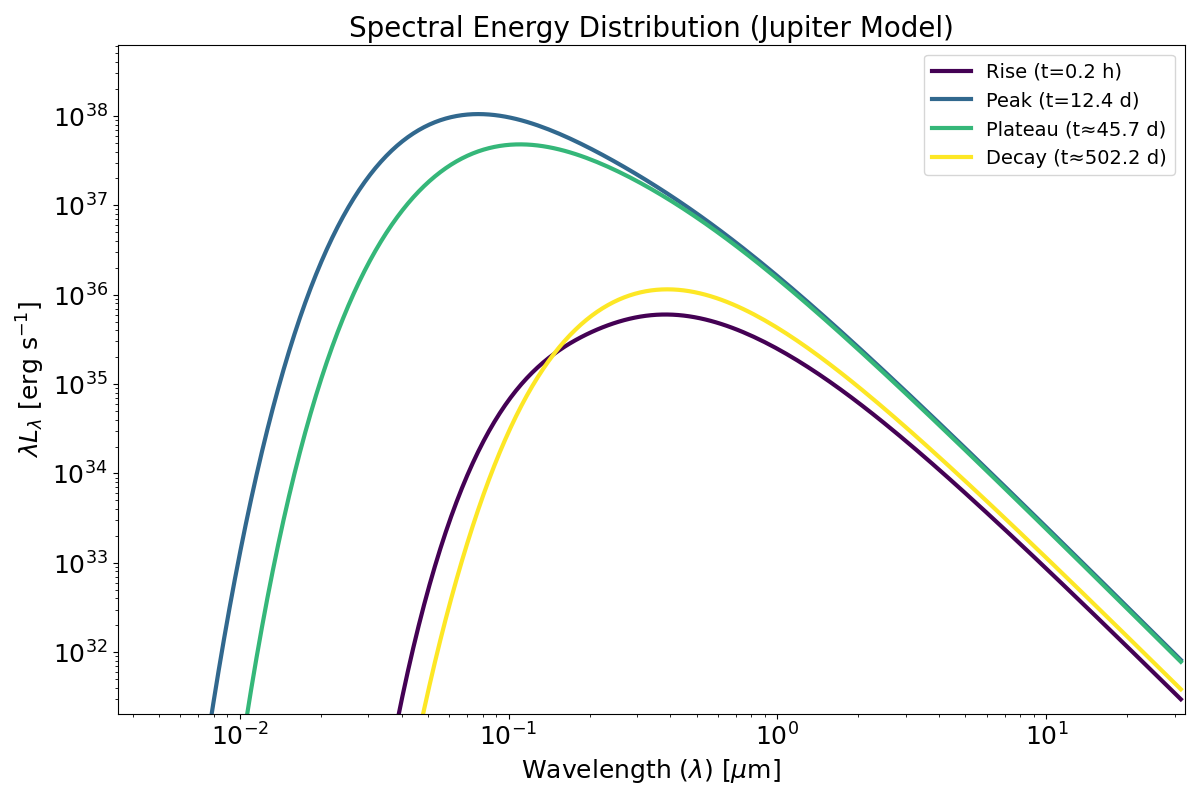}
    \caption{
    Evolution of the spectral energy distribution (SED) for the fiducial Jupiter model at four representative times: the initial rise (t=0.2 h), the bolometric peak (t=12.4 d), the subsequent plateau (t$\approx$46 d), and the late decay phase (t$\approx$502 d). The peak of the emission clearly shifts to longer, redder wavelengths as the disk cools over time.
    }
    \label{fig:sed_evolution}
\end{figure}

The thermal evolution of the disk is best understood through its intrinsic spectral energy distribution (SED), shown in Figure~\ref{fig:sed_evolution} at four key epochs. Early in the evolution, during the hot rise and peak phases, the emission is dominated by high-energy photons, with the spectral peak located in the ultraviolet at $\lambda_{\rm peak} \lesssim 0.1 \, \mu$m. As the disk expands and cools viscously through the plateau and decay phases, the SED peak shifts significantly to longer wavelengths, moving through the optical and into the near-infrared. This migration of the emission peak is the fundamental physical reason for the colour evolution seen in observations, as the system transitions from a hot, blue state to a cooler, redder one.

\subsubsection{Multi-wavelength light curves and colour evolution}

\begin{figure}
    \centering
    \includegraphics[width=0.9\linewidth]{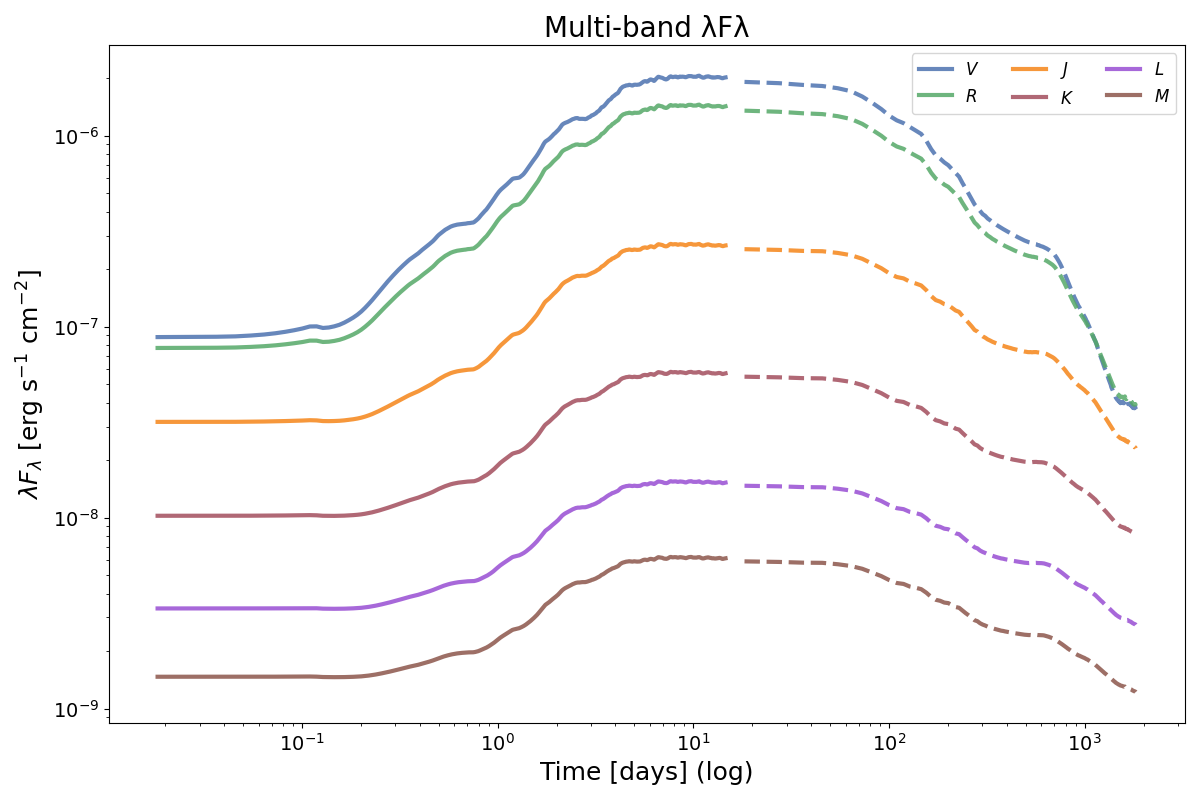}
    \includegraphics[width=0.9\linewidth]{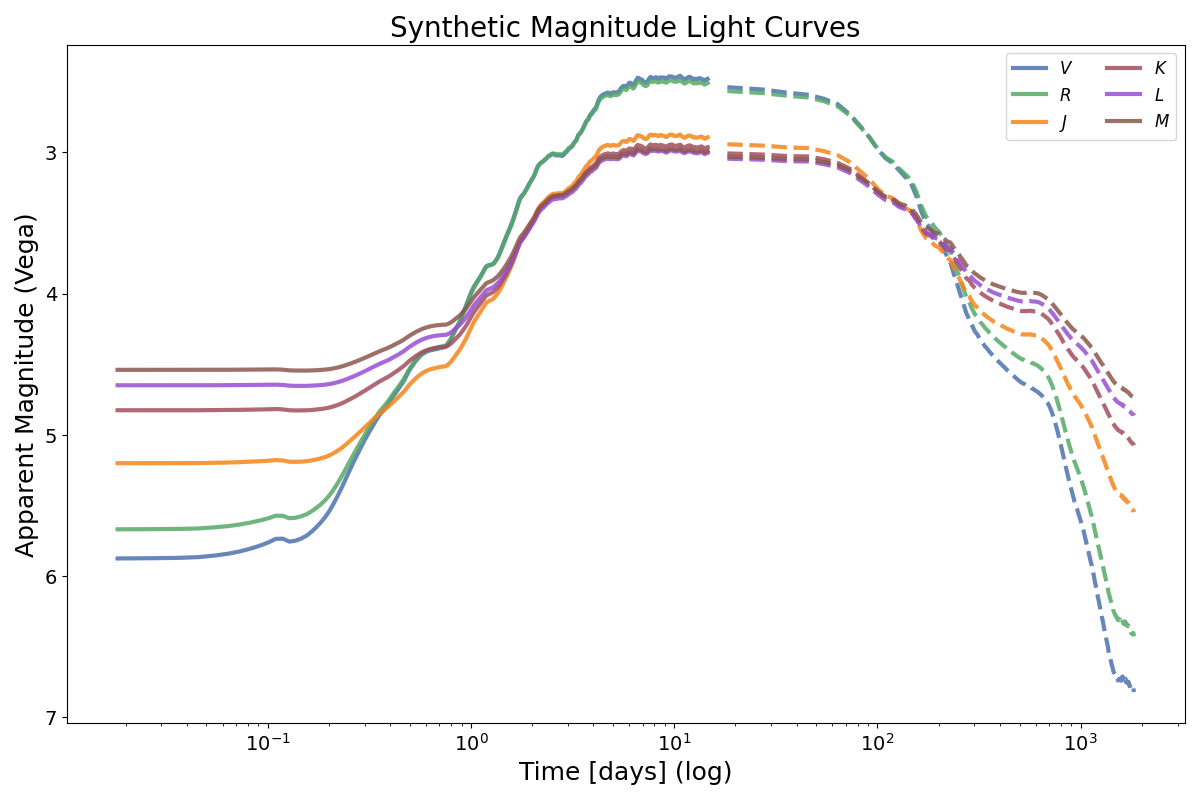}
    \includegraphics[width=0.9\linewidth]{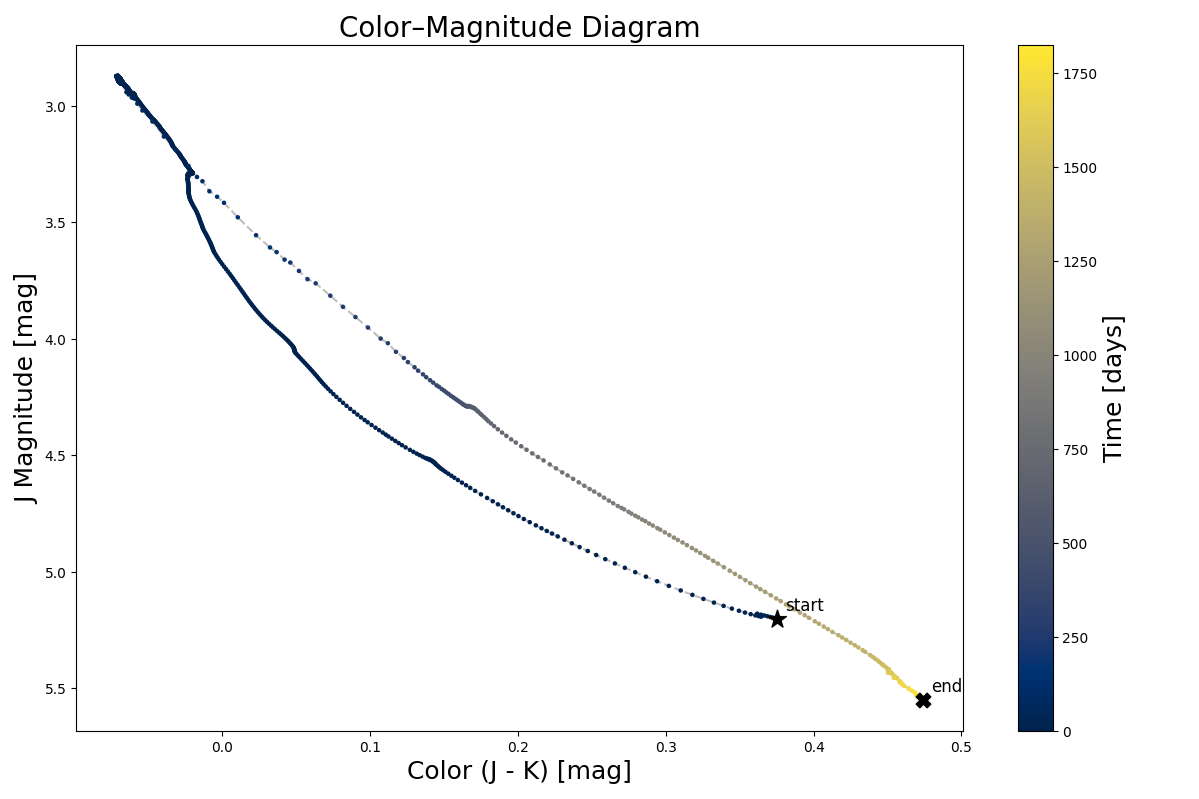}
    \caption{
        Predicted multi-wavelength signatures for the fiducial Jupiter model.
        \textbf{Top:} Multi-band light curves in flux units.
        \textbf{Middle:} Apparent Vega magnitudes.
        \textbf{Bottom:} Colour-magnitude diagram showing the "bluer-when-brighter" evolution.
    }
    \label{fig:multi_wavelength}
\end{figure}

The observable consequences of this spectral evolution are presented in Figure~\ref{fig:multi_wavelength}. The multi-band light curves (panels a and b) show that the transient is brightest at shorter, optical wavelengths near its peak, reaching an apparent magnitude of $m_V \approx 2.46$. As the disk cools, the peak of its emission shifts toward the near-infrared, causing the optical bands to fade significantly faster than the infrared ones. For instance, the J-band flux takes nearly twice as long to decay by a factor of 10 ($t_{10} \approx 1461$~d) compared to the V-band ($t_{10} \approx 767$~d).

This thermal evolution is most clearly illustrated in the colour-magnitude diagram (CMD) in panel (c). The system's track begins at a relatively faint, red state before moving upwards and to the left as it heats and brightens. This trajectory, where the object becomes bluer as it gets brighter, is a direct consequence of the SED peak shifting to shorter wavelengths with increasing temperature. Following the peak, the track reverses, moving downwards and to the right as the disk gradually cools and fades. This "bluer-when-brighter" behaviour is therefore a classic and robust signature of a transient object undergoing rapid heating followed by a prolonged cooling phase.

\section{Discussion}
\label{sec:discussion}

Our hydrodynamic simulations reveal that planetary TDEs can produce a diverse range of luminous transients. The morphology of the resulting light curve is a sensitive function of both the progenitor planet's mass and its initial orbital eccentricity. We now discuss the physical interpretation of these results and their implications for observing such events.

\subsection{Comparison with analytical estimates}

To check that our simulation results are compatible with fundamental physical principles, we compare their energetics to a simplified analytical model. Assuming the total radiated energy $\Delta E$, is released over a characteristic circularization timescale, $t_{\rm circ}$, from the surface of a newly formed disk with area $A \approx 2\pi r_p^2$ (where $r_p$ is the disruption radius), the effective temperature can be estimated as:

\begin{equation}
    T_{\rm eff} \sim \left[ \frac{\Delta E}{2\pi \sigma_B t_{\rm circ} r_p^2} \right]^{1/4}.
\end{equation}

For our fiducial Jupiter model ($e=0$), using the simulated values $\Delta E = 1.14 \times 10^{45}$~erg and $t_{\rm circ} \approx t_{\rm peak} = 12.4$~days, we obtain an analytical estimate of $T_{\rm eff} \approx 55,000$~K. For the eccentric Jupiter model ($e=0.5$), the much shorter timescale $t_{\rm circ} \approx 1.0$~day and $\Delta E = 9.32 \times 10^{44}$~erg yield $T_{\rm eff} \approx 98,500$~K. These estimates are in excellent agreement with the luminosity-weighted average temperatures found in our simulations at peak luminosity ($\langle T_{\rm eff} \rangle_L \approx 48,400$~K and $\approx 85,400$~K, respectively), see Table \ref{tab:models_comparison}. This consistency shows that the energetics and thermal properties of the simulated events are well-grounded in the principles of orbital energy dissipation. \footnote{We note that the peak luminosity in the fiducial model slightly exceeds the Eddington limit for a $1\,M_{\odot}$ star ($L_{\rm Edd} \approx 1.3 \times 10^{38}$ erg s$^{-1}$). While super-Eddington accretion has been predicted for planetary TDEs around high-mass stars \citep{Elbakyan+2024}, our 2D results demonstrate that this regime is also accessible for solar-mass hosts. Such rates could trigger vertical expansion and radiatively driven mass loss \citep[e.g.,][]{StrubbeQuataert2009}, suggesting that the system might experience a phase of enhanced outflows during the peak, although the secular viscous evolution captured in our simulations dominates the long-term behaviour.}

\subsection{The dichotomous effect of orbital eccentricity}
\label{sec:dichotomous effect}

The distinct orbital configurations explored in our simulations likely correspond to different dynamical pathways within the evolving planetary system. Our fiducial circular model ($e=0$) is consistent with the endpoint of smooth, disk-driven migration (Type II), where the planet gradually spirals inward through the gaseous disk until it reaches the tidal disruption radius. In this scenario, the planet is naturally embedded in the circumstellar medium. Conversely, the eccentric model ($e=0.5$) represents a dynamical scattering scenario—potentially driven by instabilities or planet-planet interactions in the outer regions—which flings the planet onto a plunging trajectory through the background gas. The presence of the background disk in our simulations ensures that the hydrodynamic interaction with the ambient medium is self-consistently captured alongside the tidal disruption itself.

A significant finding of this work is the dichotomous effect of initial orbital eccentricity on planets of different masses. For a Jupiter-like planet, a plunging orbit ($e=0.5$) results in a dramatically faster, hotter, and more hydrodynamically violent transient compared to a circular orbit. In stark contrast, the same eccentricity causes the disruption of a Neptune-like planet to be significantly delayed, more luminous, and more prolonged.

We propose that this divergence is rooted in the planets' internal structures. A lower-density, more compressible gas giant like Jupiter may be susceptible to a catastrophic, single-passage disruption during a high-velocity eccentric encounter. This rapid shredding and shocking of the material would lead to a fast circularization of the debris and a correspondingly rapid rise to a hot, volatile peak.

Conversely, a denser, more compact planet like Neptune may be more resilient to a single plunging encounter. The tidal forces might only strip its outer gaseous envelope initially, leaving a more intact core. The subsequent evolution would then be governed by the slower circularization of this stripped material, or potentially by multiple passages of the remnant core, leading to a delayed, broader, and more sustained light curve. If this interpretation is correct, it suggests that the morphology of a planetary TDE light curve encodes information not only about the orbital parameters but also about the internal structure of the disrupted planet itself, offering a powerful potential diagnostic for future observations of these rare events.

It is worth noting, however, that the resilience of the Neptune-like progenitor inferred from our models assumes a differentiated structure typical of core accretion formation. Recent MHD simulations of disk fragmentation suggest that Neptune-mass planets can also form via gravitational instability, potentially resulting in coreless structures \citep{Kubli+2023, Kubli+2025}. Such planets would lack the dense core that delays disruption in our fiducial model, likely leading to a faster and more complete disruption similar to the gas giant scenarios.

\subsection{Implications for observational signatures}

Our simulations show that the planetary debris does not remain confined to a compact ring at the initial disruption radius. Instead, as highlighted by the evolution at peak luminosity (t=12.4 d), the material rapidly spreads outwards, forming an extended and luminous transient accretion disk (Figure \ref{fig:disk_evolution}). This process has a direct impact on the expected observational signature.

The rapid outward transport of material heats a significantly larger surface area than a simple, static ring model would imply. As shown in the effective temperature maps, high temperatures ($>10^5$ K) are not confined to the innermost edge but are present throughout the extended disk structure. Consequently, the total luminosity of the transient is not solely produced by a small, hot central region, but includes a substantial contribution from this larger, cooler, outer area. This has two key implications for observations. First, the large emitting surface can produce a higher overall luminosity than a more compact disk. Second, it directly shapes the spectral energy distribution (SED) of the transient. The combination of a hot inner edge and a warm, extended outer region would produce a broader SED than a single-temperature blackbody, potentially making the event appear redder in some optical/IR colour indices.

Beyond the material that forms the accretion disk powering the transient, the hot and turbulent nature of the disk at peak provides favorable conditions for launching outflows or unbinding a fraction of the debris. This scenario connects directly to the secondary consequences of planetary disruptions. In particular, \citet{Sucerquia+Montesinos2025} investigated the fate of volatile-rich gas released during such events. Their work showed that this material, especially from giant planets like the one modeled here, could escape the inner system and be gravitationally captured by outer planets, forming transient secondary atmospheres. Therefore, a single planetary TDE could produce two distinct sets of observational phenomena: an immediate, bright accretion-powered transient, and a potential long-term signature of atmospheric contamination on surviving planets within the same system.

Furthermore, the relevance of forming and accreting a planetary debris disk extends beyond main-sequence stars; the process is the leading explanation for a widely observed phenomenon: the metal pollution in white dwarf atmospheres. While many polluted white dwarfs are thought to accrete tidally disrupted rocky bodies \citep{Jura+2003, Koester+2014}, the fate of gas giants in these systems is also a key question. Our simulations model the catastrophic \textit{disruption} of a gas giant, but an alternative channel for polluting a white dwarf with volatile elements has also been identified. \citet{Schreiber+2019Natur} have proposed that some pollution patterns, especially those showing volatile elements, can be explained by the accretion of gas \textit{evaporating} from the atmosphere of a close-in, intact giant planet. In their model, this evaporated material forms a gaseous disk from which the white dwarf accretes. Although the disk-forming mechanism (tidal disruption vs. evaporation) differs from our model, both scenarios involve the accretion of gas-rich material from a giant planet, reinforcing the idea that these processes are viable and have observable consequences across different stellar evolutionary stages.

\subsection{Distinguishing TDEs from FUor outbursts}

The detailed hydrodynamic and thermodynamic framework presented in this work, particularly those arising from eccentric orbits, provide a compelling alternative mechanism for a class of anomalous outbursts in Young Stellar Objects (YSOs) that defy standard classification. While luminous outbursts in YSOs are typically attributed to FUor or EXor events (for reviews, see \citealt{Hartmann+1996}, \citealt{Audard+2014}, and \citealt{Fischer+2023})—which are understood as thermal or gravitational instabilities in the \textit{innermost} disk ($< 1$~au) driving enhanced accretion onto the star—some observed events show contradictory evidence.

A prime example is VVV-WIT-13, a long-duration ($\sim$2000--3000 day) outburst interpreted by \citet{Guo+2025} as originating from a "hot ring" at a large orbital distance ($\sim 3$~au). This "outside-in" heating mechanism is inconsistent with a classical FUor event. Our simulations provide a robust physical basis for this scenario. As shown in our thermal evolution maps (Fig.~\ref{fig:disk_evolution}, Peak panel), the dissipation of orbital energy during a TDE is a global, turbulent process that heats the entire extended region where the debris circularizes, consistent with the "hot ring" model and distinct from the inner-disk heating of a FUor.

Furthermore, the light curves of standard FUor/EXor instabilities are characterized by a smooth rise followed by a decades-long, stable plateau (as seen in the classical FU Ori) or a repetitive, faster decay (as in EX Lupi) (see Fig. 4 of \citealt{Fischer+2023}). In sharp contrast, our simulations of eccentric disruptions ($e=0.5$) predict an intrinsically "highly volatile" light curve, featuring a complex, multi-peaked structure (Fig.~\ref{fig:bolometric_comparison}, right panel). This hydrodynamic instability, driven by the violent circularization of the eccentric material, is a key predictive signature of a planetary TDE. This offers a clear observational discriminant to distinguish eccentric TDEs from the smoother light curves of viscosity-driven FUor outbursts.

\subsection{Alternative Scenario: A star–planet merger}

An important alternative channel for producing a stellar transient is a direct star–planet merger, such as the event observed in ZTF SLRN-2020 \citep{De+2023}. While such events can resemble luminous red novae, our TDE models predict several distinguishing features.

First, the rise time provides a key discriminant. The extremely rapid rise of our eccentric Jupiter model ($\sim$1~day) is significantly faster than the $\sim$10--25 day plateaus typically associated with merger events powered by recombination in an outflow. While our fiducial model shows a slower rise, its subsequent evolution is governed by an accretion disk, leading to a power-law-like decay that is physically distinct from the phenomenology of a single-ejection merger.

Second, the colour evolution offers another way to differentiate. Our models universally predict a "bluer-when-brighter" behaviour, characteristic of a disk heating and then cooling. This contrasts with merger scenarios involving significant and rapid dust formation, which would cause the event to become redder as it evolves \citep[e.g.,][]{Pejcha2016, MetzgerPejcha2017}. The combination of rise time, decay shape, and colour evolution thus provides clear observational pathways to distinguish a disk-driven planetary TDE from a merger-driven transient.

\subsection{Comparison with stellar TDEs}

The late-time evolution of the accretion disk provides a powerful tool to connect the physics of planetary TDEs with the more established field of stellar TDEs by supermassive black holes (SMBHs). Hydrodynamic simulations of the disk phase in stellar TDEs have shown that, once the disk is formed and settles, its evolution is governed by the viscous diffusion of material initially supplied by the fallback of stellar debris \citep{Montesinos+Pacheco2011, Montesinos+Pacheco2013}. A cornerstone prediction of this fallback accretion model, first proposed by \citet{Rees1988}, is that the luminosity should decay as a power-law with time, following the canonical relation $L \propto t^{-5/3}$.

A remarkable result from our new simulations is the degree to which our planetary TDE models adhere to this prediction. Our fiducial Jupiter model ($e=0$), which represents the most idealized case, exhibits a late-time decay index of $\alpha_{\rm decay} \approx -1.76$. This value is in good agreement with the theoretical prediction of $-5/3 \approx -1.67$. This consistency suggests that the fundamental physics of viscous accretion in a post-disruption disk may be scalable across vastly different astrophysical regimes—from a Jupiter-mass planet orbiting a solar-mass star to a solar-mass star orbiting a million-solar-mass black hole.

Furthermore, the deviations from this canonical decay rate in our other models serve as a valuable physical diagnostic. Just as different initial conditions in stellar TDE models can produce features like "bumps" or altered decay rates \citep{Montesinos+Pacheco2011}, the diverse decay rates seen in our eccentric and Neptune models can be interpreted as direct signatures of the event's specific properties. This suggests that a detailed analysis of the late-time light curve shape offers a potential pathway to characterize the physics of the disruption process, such as the initial orbital parameters of the event and even the internal structure of the disrupted planet.

\subsection{Vertical transport and 3D effects}
\label{sec:dimensionality}

Finally, we address the limitations inherent to the 2D vertically integrated approach. While this approximation effectively captures the global viscous evolution and radial angular momentum transport, it necessarily filters out vertical dynamical effects. In a full 3D treatment, the intense heating during the peak luminosity phase could drive vertical convection and challenge the assumption of vertical hydrostatic equilibrium. Specifically, strong shocks could trigger rapid vertical expansions, creating local scale-height inhomogeneities or ``bumps'' that might render the flow unstable; these phenomena are geometrically suppressed in a thin-disk approximation. Furthermore, our finding that the peak luminosity can exceed the Eddington limit even for solar-mass hosts, a regime previously highlighted for massive stars \citep{Elbakyan+2024}, suggests that a full 3D simulation might reveal global vertical inflation or radiatively driven winds. Capturing these effects, including the potential for strong outflows that unbind debris, requires not only higher computational resources but also the implementation of explicit vertical energy advection and radiation transport routines. We therefore reserve the exploration of these three-dimensional phenomena for future work, considering the present 2D results as the necessary baseline to establish the global mass budget and temporal evolution of the debris.

\section{Conclusions}
\label{sec:conclusions}
We have performed 2D hydrodynamic simulations of planetary tidal disruption events, systematically investigating the effects of planet mass (Jupiter vs. Neptune) and initial orbital eccentricity ($e=0$ vs. $e=0.5$) on the resulting observational signatures. This work presents a predictive framework for the identification of these events in time-domain surveys. The primary conclusions of this study are as follows:

\begin{enumerate}
    \item Planetary TDEs produce luminous, long-duration transients. Our models predict peak bolometric luminosities of $L_{\rm bol} \sim 10^{38}$~erg~s$^{-1}$ and a luminous lifetime on a timescale of a few years, providing a quantitative basis for observational searches.

    \item The light curve morphology is a sensitive diagnostic of the event's initial conditions. We find a clear contrast between disruptions from circular orbits, which produce slow rises and smooth plateaus, and those from eccentric orbits. The latter, for a Jupiter-like planet, generate a distinct signature: a very fast rise to peak ($\sim$1~day) followed by a highly volatile phase.
    
    \item The effect of eccentricity is not universal and depends on the progenitor's internal structure. For a Jupiter-like planet, eccentricity accelerates the event, while for a denser Neptune-like planet, it delays and prolongs it. This suggests that a detailed analysis of the light curve shape could be used to constrain not only the orbital parameters but also the internal properties of the disrupted exoplanet.
    
    \item A robust "bluer-when-brighter" colour evolution is a common signature across all our models. This behaviour, driven by the heating and subsequent cooling of the transient accretion disk, serves as a key observational discriminant to distinguish these events from other types of transients.
    
    \item The underlying accretion physics appears scalable across different mass regimes. The late-time decay of our fiducial Jupiter model ($\alpha_{\rm decay} \approx -1.76$) is in good agreement with the canonical $L \propto t^{-5/3}$ law predicted for stellar TDEs around supermassive black holes, suggesting a common nature for viscous processes in post-disruption disks.
\end{enumerate}

These detailed predictions can be used to develop specific search criteria for planetary TDEs in current and upcoming surveys like the Vera C. Rubin Observatory's Legacy Survey of Space and Time (LSST), enabling the identification and study of the final, dynamical stages of planetary system evolution.
\begin{acknowledgements}
      We thank the anonymous referee for their constructive comments that helped to improve the quality of this paper.
      M.M. acknowledges financial support from FONDECYT Regular 1241818. A.B acknowledges support from the Deutsche Forschungsgemeinschaft (DFG, German Research Foundation) under Germany's Excellence Strategy – EXC 2094 – 390783311. This project was supported by the European Research Council (ERC) under the European Union Horizon Europe research and innovation program (grant agreement No. 101042275, project Stellar-MADE). V.E. acknowledges support by the Ministry of Science and Higher Education of the Russian Federation (State assignment in the field of scientific activity 2023, GZ0110/23-10-IF).
\end{acknowledgements}

\bibliographystyle{aa}
\bibliography{astro}

\end{document}